\documentclass{aa}  
\usepackage{hyperref}
\usepackage{graphicx}
\usepackage{txfonts}
\begin{document}

   \title{Enhanced AGN Activity in Overdense Galactic Environments at $2 < z < 4$}


   \author{Ekta A. Shah\inst{1},
          Brian C. Lemaux\inst{2,1},
          Benjamin Forrest\inst{1},
          Nimish Hathi\inst{3},
          Lu Shen\inst{4},
          Olga Cucciati\inst{5},
          Denise Hung\inst{2},
          Finn Giddings\inst{6},
          Derek Sikorski\inst{6},
          Lori Lubin\inst{1},
          Roy R. Gal\inst{6},
          Giovanni Zamorani\inst{5},
          Emmet Golden-Marx\inst{7,8},
          Sandro Bardelli\inst{5},
          Letizia Pasqua Cassara\inst{9},
          Bianca Garilli\inst{9},
          Gayathri Gururajan\inst{10,5,11,12},
          Hyewon Suh\inst{2},
          Daniela Vergani\inst{5},
          Elena Zucca\inst{5}
          }

   \institute{Department of Physics and Astronomy, University of California, Davis, One Shields Avenue, Davis, CA, 95616, USA\\
              \email{eashah@ucdavis.edu}
         \and
            Gemini Observatory, 670 N. A’ohoku Place, Hilo, Hawai\textquotesingle i, 96720, USA
        \and
        Space Telescope Science Institute, Baltimore, MD 21218, USA
        \and Department of Physics and Astronomy, Texas A\&M University, College Station, TX, 77843-4242, USA
        \and INAF-Osservatorio di Astrofisica e Scienza dello Spazio, Via Gobetti 93/3, I-40129, Bologna, Italy
        \and University of Hawai\textquotesingle i, Institute for Astronomy, 2680 Woodlawn Drive, Honolulu, HI 96822, USA
        \and
        Department of Astronomy, Tsinghua University, Beijing 100084, China
        \and
        INAF - Osservatorio astronomico di Padova, Vicolo Osservatorio 5, 35122 Padova, Italy
        \and
        INAF-IASF Milano, Via Alfonso Corti 12, 20159 Milano, Italy
        \and
        University of Bologna – Department of Physics and Astronomy “Augusto Righi” (DIFA), Via Gobetti 93/2, 40129 Bologna, Italy
        \and 
        SISSA, Via Bonomea 265, I-34136 Trieste, Italy
        \and
        IFPU - Institute for fundamental physics of the Universe, Via Beirut 2, 34014 Trieste, Italy
        \\
             }

   \date{Received xx; accepted xx}

   \abstract
{We conduct a study on the relationship between galaxy environments and their active galactic nuclei (AGN) activity at high redshifts ($2.0<z<4.0$). Specifically, we study the AGN fraction in galaxies residing in a range of environments at these redshifts, from field galaxies to the densest regions of highly overdense peaks in the GOODS-S extragalactic field. Utilizing the extensive photometric and spectroscopic observations in this field, we measure local- and global-overdensities over a large a range of environments, including in massive (M$_{tot}$$\geq10^{14.8}$M$_\odot$) protostructures reported in \citet{shah24}. We employ a multi-wavelength AGN catalog \citep{lyu2022}, consisting of AGN in nine different categories. Our analysis shows a higher AGN fraction (10.9$^{+3.6}_{-2.3}$\%) for galaxies in the highest local-overdensity regions compared to the AGN fraction (1.9$^{+0.4}_{-0.3}$\%) in the corresponding coeval-field galaxies at $2.0<z<4.0$ (a $\sim4\sigma$ difference). This trend of increasing AGN fraction in denser environments relative to the field is present in all redshift bins. We also find this trend consistently present in all five AGN categories that have a sufficient number of AGN to make a meaningful comparison: mid-IR SED, mid-IR color, X-ray luminosity, X-ray-luminosity-to-radio-luminosity-ratio, and optical-spectroscopy at $2.0<z<4.0$. Our results also demonstrate a clear trend of higher ($\sim4\times$) AGN fractions in denser local overdensity environments for a given stellar mass. Additionally, we observe the same trend (though at a lower significance) with the global environment of galaxies, measured using a metric based on the projected distance of galaxies from their nearest massive ($M_{tot}>10^{12.8}M_\odot$) overdense ($\sigma_\delta>5.0$) peak, normalized with respect to the size of the peak. These findings indicate that the prevalence of AGN activity is highly dependent on the environment in which a host galaxy resides, even at early times in the formation history of the Universe.}

\titlerunning{AGN Activity in Different Environments at $2<z<4$}
\authorrunning{E.~A.\ Shah et al.}
\maketitle
  
%

\section{Introduction}

Galaxies reside in a variety of environments - ranging from sparsely populated field regions where galaxies have few neighbors, to dense galaxy clusters in which thousands of galaxies can reside in a relatively small region. The environment of a galaxy plays a crucial role in its formation and evolution. The gravitational potential well of large dense structures affect the motion of galaxies as well as their interaction and merger rate with other galaxies \citep{mihos2003, kim2024}. For these reasons among many others, large dense structures can significantly alter the properties of their constituent galaxies such as morphology, stellar mass, star formation rate (SFR), and active galactic nuclei (AGN) activity \citep{dressler1980, balogh2000,edler2024,rihtar2024}, as well as many other properties.

The supermassive black hole at the center of a galaxy that is actively accreting matter and emitting intense radiation is called an `AGN'. As more material gets accreted onto supermassive black holes, they grow, eventually leading to triggering of the AGN \citep[e.g.,][]{silk1998,fabian1999}. The feedback caused by AGN emission can affect the properties of galaxies such as its star formation \citep[e.g.,][]{silk1998, dimatteo2005, belli2023}, morphology \citep[e.g.,][]{dubois2016}, and chemical enrichment of the intragalactic medium \citep[e.g.,][]{hamann1999,bower2008, kara2024} among many other properties. Therefore, the interplay between black hole growth and AGN feedback is a critical aspect of galaxy evolution as it connects the growth of the black hole to the properties and evolution of the host galaxy. 

AGN are typically classified into various categories based on their observed properties. For example, radio-loud AGN emit strong radio waves relative to the strength of emission in other bandpasses (typically optical), likely due to the presence of relativistic jets powered by the central supermassive black hole.  Radio-quiet AGN have weaker radio emission, possibly due to the absence or weakness of these jets \citep[e.g.,][]{urry1995, wilson1995}. Based on optical spectra, AGN with broad emission lines, known as Type 1, are thought to provide a direct view of the central engine of the AGN and broad-line region, whereas Type 2 AGN, which exhibit narrow emission lines, are thought to be viewed through an optically thick dusty torus that obscures the central engine and broad-line region \citep[e.g.,][]{almeida2011, villaroel2014}. The re-radiation of absorbed light from dust surrounding the AGN can cause an IR-bright AGN \citep[e.g.,][]{stern2005,donley2012}. Some types of AGN can also be identified by their strong X-ray emission \citep[e.g.,][]{rees1984, ueda2003, buchner2014}. Among the most extreme examples of AGN are quasars, which are so luminous that they often outshine their host galaxies \citep[e.g.,][]{antonucci1993, urry1995, silk1998}. Additionally, the ``little red dots'' observed by the James Webb Space Telescope (JWST) could represent a population of heavily obscured, high-redshift AGN, providing insights into early black hole growth \citep[e.g.,][]{matthee2024}. Different processes may be responsible for triggering different types of AGN. Since an AGN identified using one criterion does not necessarily meet the criteria in all other categories simultaneously, the results of AGN studies can vary depending on the type of AGN used.

The complex connection of AGN activity and dense environments of galaxy clusters has been studied in-depth at low-to-intermediate redshifts ($z\lesssim1.5$). 
In the local Universe, observational studies show that luminous AGN are fractionally less common among cluster galaxies compared to coeval field galaxies \citep{kauffman2004,popesso2004}, while the population of less-luminous AGN does not show a significant difference with environment \citep{martini2002,miller2003,martini2006,haggard2010}. \citet{best2004} find that AGN are preferentially found in large-scale environment (such as moderate groups and clusters) compared to the local environment of the galaxy. At $0.65<z<0.96$, \citet{shen2017} show that radio-AGN are more likely to be found in dense local environments of the large-scale structures (LSSs) identified as a part of the Observations of Redshift Evolution in the Large-Scale Environments (ORELSE) survey \citep{lubin2009}. However, at $1<z<1.5$, \citet{martini2013} show consistent X-ray and MIR-selected AGN fractions in the clusters and field. Some studies show significant decrement in the fraction of X-ray AGN with decreasing cluster-centric radius, from $r_{500}$ to central regions \citep[e.g.,][]{ehlert2013, pentericci2013, koulouridis2024}. Additionally, at $0.65<z<1.28$, \citet{rambaugh2017} generally do not observe any strong connection between AGN activity and location within the LSSs in the ORELSE survey. 
However, they observed a notable exception in two dynamically most unrelaxed LSSs (SG0023 and SC1604), where there was an overabundance of kinematic pairs, suggesting that merging activity could be a significant factor in AGN triggering. This highlights that at least some LSS environments might play a role in triggering AGN activity through merging processes at these redshifts. On the other hand, studies such as \citet{klesman2012} do not see any significant correlations between AGN activity and cluster properties such as mass, X-ray luminosity, size, morphology, and redshift at $0.5<z<0.9$ in their sample of AGN identified based on optical variability, X-ray emission, or mid-IR power-law spectral energy distributions (SEDs). 

The complexity of the relation between dense environments of galaxies and their AGN activity is in some ways heightened at high redshift. Some studies on individual protoclusters, such as \citet{krishnan2017} (X-ray AGN; $z\sim1.7$), \citet{tozzi2022} (X-ray AGN; $z\sim2.156$), \citet{digby2010} (AGN identification using emission-lines in optical/near-IR spectra; $z\sim2.2$), and \citet{monson2023} (X-ray AGN; $z\sim3.09$), all of which show enhanced AGN activity in protoclusters compared to field. \citet{lemaux2014} show a strong AGN actvity in the brightest protocluster galaxy of Cl J0227-0421 at $z\sim3.3$. However, other studies, such as \citet{macuga2019} (X-ray AGN; $z\sim2.53$), do not show this same trend. Similarly, \citet{lemaux2018} do not see any AGN activity in member galaxies of a massive protocluster at $z\sim4.57$. A large sample of protostructures\footnote{We use the more agnostic term  ``protostructures'' (instead of protoclusters) throughout the paper as we are unsure of the fate of the systems reported in this paper.} over a large redshift range is required to understand the overall nature of this trend. Furthermore, as highlighted in \citet{hashiguchi2023}, populations of various types of AGN should be used to study the environment-AGN connection. Each type of AGN provides unique insights into different aspects of AGN activity, revealing information such as the central engine, obscuring structures, and the interaction of an AGN with its surroundings. A diverse approach encapsulating different types of AGN is key for capturing a broad spectrum of AGN phenomena across various redshifts. However, achieving this across a large redshift range requires deep, multi-wavelength observations, which are challenging due to the faintness of distant sources and the extensive multi-wavelength observational resources needed. These requirements make studying the environment-AGN connection at high redshift challenging. 

Utilizing the vast amount of deep multi-wavelength data in the Great Observatories Origins Deep Survey - South (GOODS-S) extragalactic field \citep{giavalisco2004,grogin2011}, we study the connection of environment and AGN activity of galaxies at $2<z<4$. We use a large sample of spectroscopically and photometrically selected galaxies spanning a wide range of environments, from field to protostructures, with the latter including five of the spectroscopically confirmed massive protostructures presented in \citet{shah24}, as well as a large number of other massive protostructures. We use a novel environmental measurement technique allowing us to define a coeval sample of field galaxies along with the samples of galaxies in denser environments. For the AGN identification, we employ the AGN catalog presented in \citet{lyu2022}, which uses deep multi-wavelength observations from X-ray to radio to provide AGN samples in nine different categories. These samples of galaxies and AGN enable us to conduct an unprecedented study on the environment-AGN connection of galaxies during an important epoch ($2<z<4$) of the cosmic history.


The layout of this paper is as follows: we describe the data and methods used for generating the galaxy and AGN samples in \S\ref{sec:data_and_method}. In \S\ref{sec:agn_frac_analysis}, we present our analysis on the change in AGN fraction with environment overdensity. We discuss our results and compare them with other studies in \S\ref{sec:discussion}. Lastly, we summarize our results in \S\ref{sec:summary}. 

\section{Data and Methods} \label{sec:data_and_method}

In this section, we first provide the details of the observations and methods used to generate the galaxy sample (\S\ref{subsec:gal_sample}) and the AGN sample (\S\ref{subsec:AGN_sample}). We then describe the spectral energy distribution (SED) fitting process used to estimate the stellar masses of galaxies in \S\ref{subsec:gal_sed_fitting} and the Voronoi Monte Carlo mapping method used to measure the local overdensity of galaxies in \S\ref{subsec:gal_env_meas}.

\subsection{Galaxy sample}\label{subsec:gal_sample}

In this study, we utilize the extensive photometric (\S\ref{gal_photometric_cat}) and spectroscopic (\S\ref{gal_spec}) observations in the Extended Chandra Deep Field South (ECDFS) field \citep{lehmer2005}, which encapsulated the GOODS-S field. The ECDFS field has deep multi-wavelength observations available across the electromagnetic spectrum \citep[e.g.,][]{zheng2004, wuyts2008, cardamone2010, dahlen2013,hsu2014} enabling a wide range of extragalactic studies \citep[e.g.,][]{kaviraj2008,lefevre2015,marchi2018,birkin2021}.

\subsubsection{Photometric catalog} \label{gal_photometric_cat}

To generate the initial galaxy sample, we select galaxies from the \citet{cardamone2010} photometric catalog for galaxies in the ECDFS field. This catalog includes deep optical medium-band photometry (18 bands) from the Subaru telescope, UBVRIz photometry from the Garching-Bonn Deep Survey \citep[GaBoDS;][]{hildebrandt2006} and the Multiwavelength Survey by Yale-Chile \citep[MUSYC;][]{gawiser2006} survey, deep near-infrared (NIR) imaging in JHK from MUSYC \citep{moy2003}, as well as \textit{Spitzer} Infrared Array Camera (IRAC) photometry obtained using the Spitzer IRAC/MUSYC Public Legacy Survey in ECDFS \citep[SIMPLE;][]{damen2011}. The multi-wavelength observations used for generating the galaxy catalogs, estimating the stellar masses of galaxies using spectral energy distribution, and utilizing the Voronoi Monte Carlo (VMC) mapping technique to calculate the overdensity values for galaxies are described in detail in \citet{shah24}. 

For this study, we only select galaxies which have their IRAC1 (3.6\thinspace $\mu$m band) or IRAC2 (4.5\thinspace $\mu$m band)  magnitudes brighter than 24.8. The selection of this cutoff has been determined by considering the 3$\sigma$ limiting depth of the IRAC images within the ECDFS field. This cut ensures a reliable detection in the rest-frame optical wavelengths at these redshifts, which helps constrain the Balmer/4000\AA\ break for galaxies at $2<z<5$ (see Lemaux et al. 2018 for details). There are 55,147 unique objects in our sample satisfying this IRAC magnitude cut. Following the method of \citet{lemaux2018}, this IRAC cut effectively results in a sample with an 80\% stellar mass completeness limit of log(M$*$/M$\odot$) $\sim$ 8.8-9.14, depending on the redshift in the redshift range $2.0<z<4.0$.

\subsubsection{Spectroscopy} \label{gal_spec}

For this study, we utilize a combination of publicly available and proprietary spectroscopic observations. We use a compilation of publicly available redshifts in the GOODS-S field compiled by one of the authors (NPH). This compilation includes spectroscopic redshifts from surveys such as VIsible Multi-Object Spectrograph \citep[VIMOS;][]{lefevre2003} VLT Deep Survey \citep[VVDS;][]{lefevre2004,lefevre2013}, the MOSFIRE Deep Evolution Field (MOSDEF) survey \citep{kriek2015}, the 3D-HST survey \citep{momcheva2016}, a deep VIMOS survey of
the CANDELS CDFS and UDS field \citep[VANDELS;][]{mclure2018,pentericci2018}, along numerous other surveys. Additionally, we also use spectroscopic redshifts from the VIMOS Ultra-Deep Survey \citep[VUDS;][]{lefevre2015}. These surveys predominantly focus on star-forming galaxies (SFGs) with $\sim 0.3- 3L^*_{UV}$ luminosity, where $L^*_{UV}$ is the characteristic UV luminosity at a given redshift \citep{schechter1976}. These galaxies are generally representative of SFGs at their given redshifts (see Lemaux et al. 2022 for details). 

Along with the publicly available spectroscopic redshifts, we also utilize our proprietary spectroscopic redshifts obtained using Keck/DEep Imaging Multi-Object Spectrograph (DEIMOS, \citealt{faber2003}) and Keck/Multi-Object Spectrometer for Infra-Red Exploration (MOSFIRE, \citealt{mclean2010,mclean2012}) as a part of the Charting Cluster Construction with VUDS and ORELSE (C3VO) survey \citep{shen2021,lemaux2022}. These observations consists of a total of 29 and 26 secure (i.e., reliability of $\gtrsim95$\%) spectroscopic redshifts obtained from five MOSFIRE masks and two DEIMOS masks, respectively. Most of these redshifts are in the range of $2.5<z<4.0$. These masks were designed to target a suspected protostructure at $z\sim3.5$ \citep[e.g.,][]{ginolfi2017,forrest2017}, which is now spectroscopically confirmed \citep[see protostructure 4: Smruti in][]{shah24}. The details of these MOSFIRE and DEIMOS observations, data reduction, the procedure for redshift estimation are described in detail in \citet{shah24}.

We combine these C3VO spectroscopic observations with the publicly available spectroscopic redshifts described above. To get the spectroscopic redshift for a given photometric object, we perform a nearest-neighbor matching within an aperture of 1$\arcsec$ centered on the coordinates of each source in the spectral catalog. For cases where multiple spectroscopic redshifts exist for a specific photometric object, we choose the most reliable $z_{spec}$, considering factors like the quality of the redshift, the type of instrument used, the integration time of the survey, and the photometric redshift. Out of the 55,147 galaxies satisfying the IRAC cut, 9111 have a secured spectroscopic redshift.

\subsection{AGN sample}\label{subsec:AGN_sample}

\begin{figure}
\includegraphics[scale=0.53, trim={0cm 0.4cm 0cm 0cm}, clip]{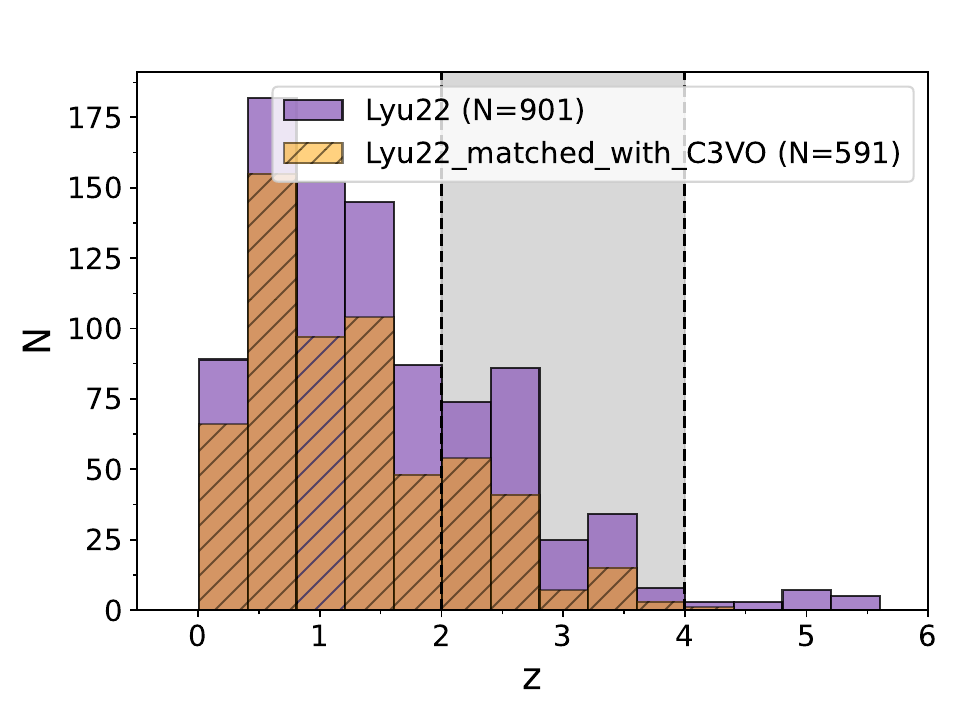}
    \caption{Redshift distribution of all 901 AGN in the \citet{lyu2022} catalog (purple histogram). This large AGN dataset includes all nine categories of AGN defined in \citet{lyu2022}. Note that the categories are not mutually exclusive, i.e., a given AGN can satisfy the selection criteria for multiple categories of AGN. The redshift values for these AGN are sourced from the \citet{lyu2022} catalog. We also overplot the redshift distribution of 591 \citet{lyu2022} AGN that are matched with the C3VO (MUSYC) sources within a projected separation of 0.5'' (orange histograms with purple lines). The redshift distribution peaks around $z\sim0.75$, and almost vanishes at $z>4$. In this study, we restrict our analysis to the redshift range of $2.0<z<4.0$ as shown by the gray shaded region between the two black dashed vertical lines.} 
    \label{fig:zhist_agn_lyu}
\end{figure}

\begin{figure}
\includegraphics[scale=0.53, trim={0cm 0.4cm 0cm 0cm}, clip]{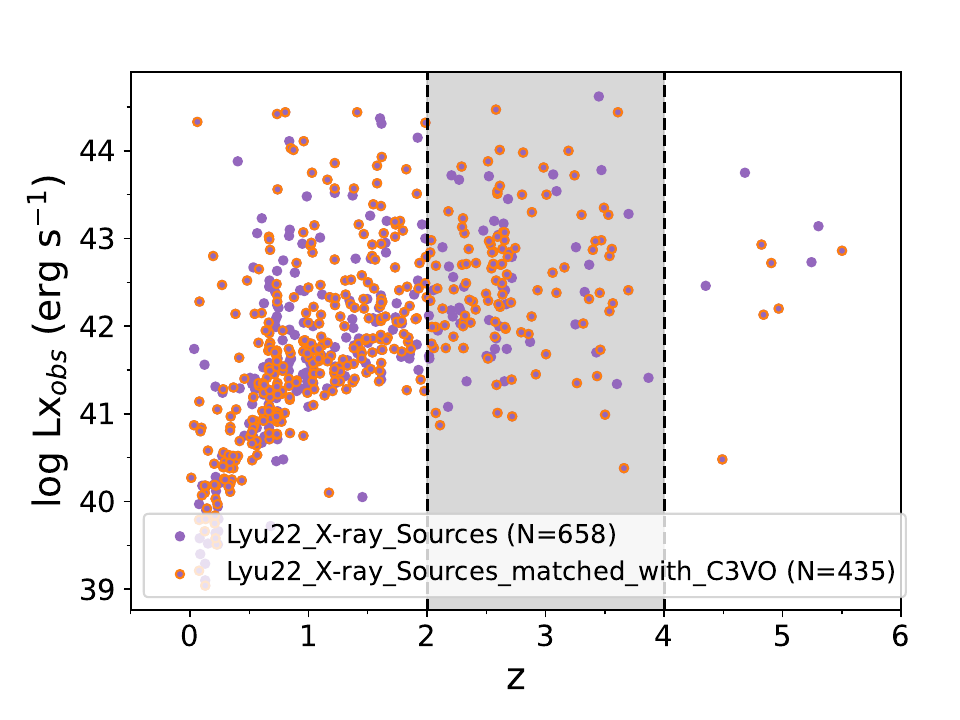}
    \caption{Distribution of the observed X-ray luminosity ($Lx_{obs}$) vs. redshift (z) for all 658 sources in \citet{lyu2022} with $Lx_{obs}>0.0$ (purple filled circles). Out of them, the 435 sources that are matched to the C3VO sources within a separation of 0.5'' are highlighted (purple filled circles with orange edge colors). The redshifts for the \citet{lyu2022} sources are from their AGN catalog, which are sourced from the 3D-HST catalog. For this study, we only focus on redshift range of $2.0<z<4.0$ as shown by the gray shaded region between the two black dashed vertical lines.} 
    \label{fig:sc_lxobs_z_lyu}
\end{figure}

In this study, we utilize the extensive AGN catalog provided by \citet{lyu2022}, which presents a comprehensive census of AGN identified by multi-wavelength observations (\S\ref{agn_obs}) from X-ray to radio in the GOODS-S/HUDF region. \citet{lyu2022} probe diverse populations of obscured as well as unobscured AGN by selecting AGN using nine different criteria (\S\ref{agn_sel_crit}) based on X-ray properties, ultraviolet to mid-infrared SEDs, optical spectral features, mid-infrared colors, radio loudness and spectral slope, and AGN variability. Using these techniques, \citet{lyu2022} generated a comprehensive sample of 901 AGN within the $\sim 170$ arcmin$^2$ 3D-HST GOODS-S footprint, which significantly expanded the number of known AGN in the region. Their analysis shows the complexity of the AGN population and indicates that no single selection method can exhaustively identify all types of AGN within a field. Hence, their extensive AGN sample identified using various selection methods is invaluable for our study, enabling our detailed analyses of how the overall AGN population as well as various types of AGN get affected by dense environments.

We note that while our density mapping (see \S\ref{subsec:gal_env_meas}) covers the larger ECDFS region compared to GOODS-S, the \citet{lyu2022} AGN sample is only available for the GOODS-S region within ECDFS. Therefore, for the AGN fraction analysis, we select galaxies exclusively from the GOODS-S region. However, the overdensity values for these galaxies in the GOODS-S field are still sourced from our density maps covering the entire ECDFS field.

\subsubsection{Observations used for generating the AGN sample}\label{agn_obs}

For generating their AGN sample, \citet{lyu2022} compiled a sample of 1886 unique GOODS-S 3D-HST (v4.1) \citep{brammer2012,skelton2014} sources with multi-wavelength counterparts identified as (i) ``radio-detected" sources in VLA 3 GHz images, (ii) ``X-ray detected sources" from the Chandra 7 Ms X-ray source catalog \citep{luo2017}, (iii) ``mid-IR detected" sources from the Spitzer MIPS 24 $\mu$m \citep{perez2008,barro2011a,barro2011b} or IRS 16 $\mu$m images or sources \citep{teplitz2011} with AGN-like IRAC colors in mid-IR, or (iv) time variable sources in the HST or Chandra images. For the radio sources, the optical to mid-IR counterparts are matched within a radius of 0.5''. For the X-ray sources, they used a matching radius of 2.0'' to identify optical to mid-IR counterparts. The \citet{lyu2022} catalog also includes redshifts of AGN sourced from the 3D-HST catalog, with spectroscopic redshifts for some objects and photometric redshifts for others. This redshift distribution of all AGN listed in the \citet{lyu2022} catalog, as well as their AGN that have a match in our C3VO catalog (see \S\ref{subsubsec:agn_matching}) are shown in Figure~\ref{fig:zhist_agn_lyu}.

\begin{figure*}
\includegraphics[scale=0.8]{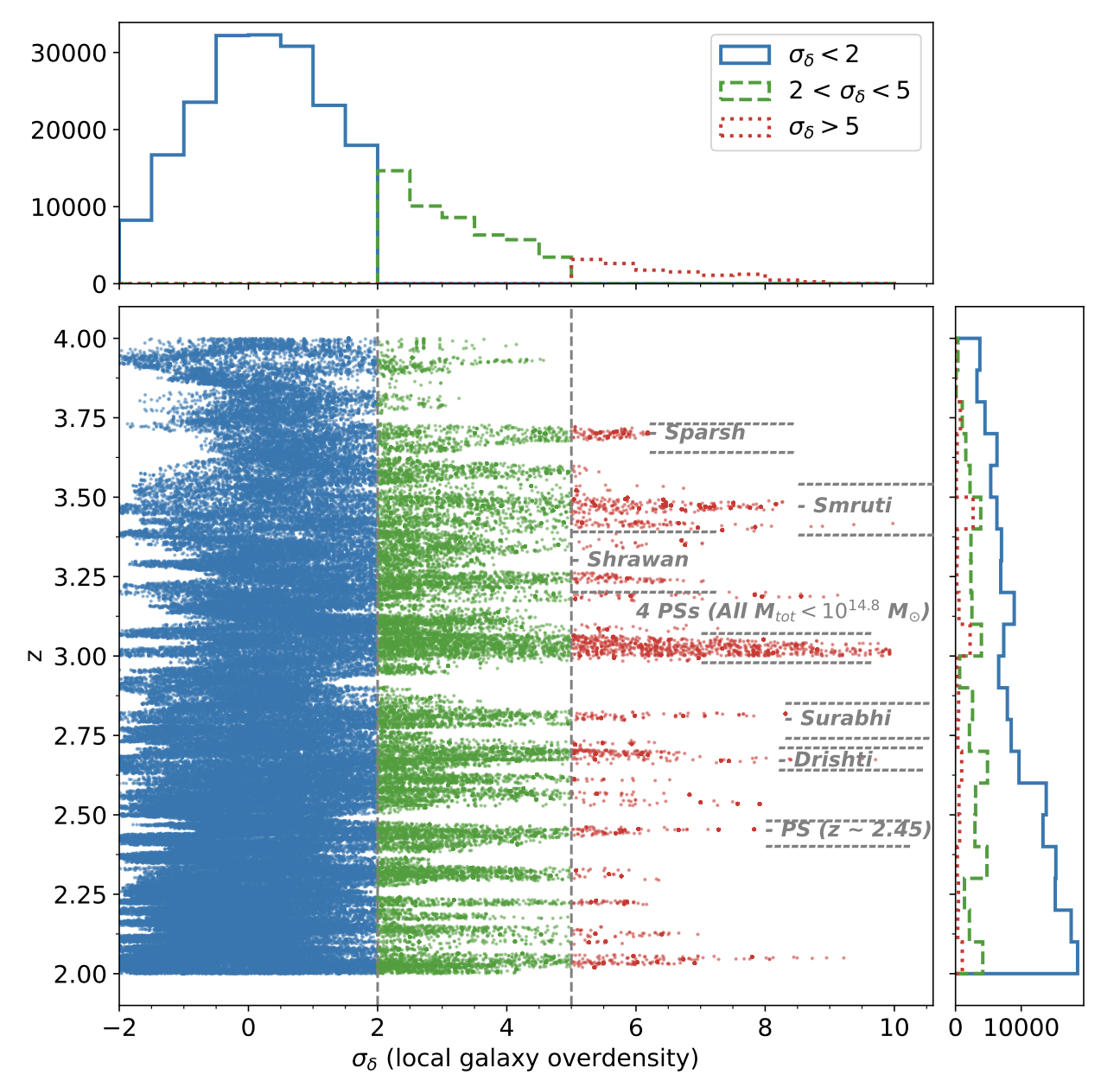}
    \caption{Distribution of redshift vs. $\sigma_{\delta}$ (local galaxy overdensity) for 120,525 objects with $2.0 < z_{gal} < 4.0$ across the 100 Monte Carlo realizations. The number of objects is approximately 100 times higher than the true number of galaxies, averaging around 1205 objects per Monte Carlo iteration. The scatter points represent galaxies with their MC realization-specific local overdensity ($\sigma_\delta$): coeval field ($\sigma_\delta<2.0$) (blue points), intermediate overdensities ($2.0<\sigma_\delta<5.0$) (green points), and highly overdense peaks  ($\sigma_\delta>5.0$) (red points). The histograms on top show the $\sigma_\delta$ distribution and the histograms on right show the redshift ($z_{gal}$) for the three $\sigma_\delta$  bins: $\sigma_\delta<2.0$ (solid blue line), $2.0<\sigma_\delta<5.0$ (dashed green line), and $\sigma_\delta>5.0$ (dotted red line). The dashed gray lines show the division of the local overdensity in the three different bins used in our analyses. We also write the given names (gray text) of all five spectroscopically confirmed massive ($M_{tot}\ge 10^{14.8} M_\odot$) protostructures that are at $z<4$ as presented in \citet{shah24} along with their redshift extent (horizontal dashed lines above and below the protostructure names). Furthermore, we also show the redshift of the most massive ($M_{tot}=10^{14.7}M_\odot$) protostructure in $2.0<z<4.0$, which is located at $z\sim2.45$, and spread over $2.40<z<2.48$. The big horizontal stripe at $z\sim3.05$ consists of 4 protostructures that are individually less massive than the protostructures reported in \citet{shah24}. Similarly, the horizontal stripe at $z\sim2.05$ consists of a protostructure that is less massive than the $z\sim2.45$ protostructure reported in this study.}
    \label{fig:z_sig_scat}
\end{figure*}

\subsubsection{AGN Selection criteria}\label{agn_sel_crit}

To probe populations of different types of AGN, \citet{lyu2022} apply various selection criteria on the multi-wavelength source catalog described above. The nine categories of AGN presented in \citet{lyu2022}, along with their selection criteria as well as the average number of AGN in our sample based on our statistical framework (see \S\ref{subsec:gal_env_meas}) in the redshift range of $2.0<z<4.0$, are briefly described below:

\begin{enumerate}
    \item \textbf{mid-IR-SED:} An AGN in this category has a notable excess emission in near-mid IR ($\lambda \sim 2-8\thinspace \mu$m) from hot- to warm-dust. There are 288 mid-IR-SED AGN in the \citet{lyu2022} catalog. \citet{lyu2022} conducted SED fitting of galaxies using the SED fitting tool \textit{Prospector} \citep{johnson2021}. They utilized the Flexible Stellar Population Synthesis (FSPS) stellar model from \textit{Prospector}, along with their own (semi) empirical models of dust emission and AGN continuum emission \citep{lyu2017a,lyu2017b,lyu2018}. To identify AGN based on SED analysis, they require that the best-SED fit contains a significant AGN component with $L_{AGN}>10^8 L_\odot$. For these cases, the SED fits are visually inspected to discard the cases for which the fitting solutions are degenerate with multiple peaks of the $L_{AGN}$ posterior as well as the cases for which the photometric constraints are too limited to determine if the AGN component exists. In cases where the classification of objects is unclear, the method involves conducting SED fitting both with and without an AGN component. If the inclusion of the AGN component results in the chi-square values being reduced by a factor of 2.5 or more, the object is then categorized as a mid-infrared SED AGN. For cases with $L_{AGN}>10^8 L_\odot$ but visual inspection of borderline or marginal, they conduct SED fitting both with and without an AGN component. Similar to other objects, for these cases, if the inclusion of the AGN component results in the chi-square values of the data being reduced by a factor of 2.5 or more, the object is categorized as a mid-infrared SED AGN. On average, there are $\sim27$ mid-IR-SED AGN in our sample.
    
    \item \textbf{mid-IR-COLOR:} These objects have mid-IR colors that are dominated by AGN warm dust emission. They are selected using the criterion of log($S_{5.8}/S_{3.6})>0.08$ and log($S_{8.0}/S_{4.5})>0.15$ from  \citet{kirkpatrick2017}. \citet{lyu2022} list 104 mid-IR-COLOR AGN in their catalog. Our AGN sample contains an average of $\sim$ 13 mid-IR-COLOR AGN.

    \item \textbf{X-ray-lum:} These objects have intrinsic X-ray luminosity higher than expected for stellar processes in galaxies. These AGN are identified using $L_{X,int}>10^{42.5}$\thinspace erg s$^{-1}$ \citep{luo2017}. For cases with slightly lower intrinsic X-ray luminosity cut ($L_{X,int}>10^{42.0}$\thinspace erg s$^{-1}$), if their SED fittings suggest low star formation rates ($\lesssim 10 M_\odot yr^{-1}$), they are selected as an AGN \citep[e.g.,][]{pereiera2011,mineo2014,symeonidis2014,algera2020}. The observed X-ray luminosity distribution with redshift is shown for all \citet{lyu2022} AGN that have $Lx_{obs} > 0$ in Figure~\ref{fig:sc_lxobs_z_lyu}. The figure also shows this  distribution for such sources that are also matched to the C3VO (MUSYC) sources within a projected separation of 0.5''. \citet{lyu2022} identified 321 AGN satisfying their X-ray-lum criteria. On average, there are $\sim$ 41 AGN identified based on X-ray-lum in our AGN sample. 
    
    \item \textbf{X2R:} These objects have X-ray (0.5–7 keV) to radio (3\thinspace GHz) luminosity ratio higher than expected for stellar processes in a galaxy. These AGN are identified using the criterion of $L_{X,int}$[erg/s]/L$_{3GHz}$[W/Hz] $> 8\times10^{18}$. In their catalog, \citet{lyu2022} report 588 X2R AGN. Our AGN sample consists of an average $\sim51$ X2R AGN. 
    
    \item \textbf{radio-loud:} These objects are radio-loud AGN with excess emission in the radio band compared with the prediction of templates of normal star-forming galaxies (SFGs). They are selected using $q_{24,obs} = log(S_{24\mu m,obs}/S_{1.4GHz,obs})$, $q_{24,obs}<q_{24,temp}$. Here, $S_{24\mu m,obs}$ and $S_{1.4GHz,obs}$ are the observed MIPS 24\thinspace$\mu$m flux and  the radio flux density at 1.4\thinspace GHz, respectively. $S_{1.4GHz,obs}$ is computed by extrapolating the 3 GHz flux assuming that the radio spectrum can be described as a power law with $\alpha = -0.7$. $q_{24,obs}$ is computed following \citet{alberts2020}. $q_{24,obs}$ is then compared to $q_{24,temp}$, which is from the radio–infrared correlations based on the \citet{rieke2009} SFG templates at the appropriate redshifts. An object that is 0.5 dex below the midpoint of the radio–infrared relation with more than $2\sigma$ significance is classified as an AGN. We note that as they compare radio emission with emission at 24 $\mu$m, their definition of radio-loud AGN differs from its canonical definition in the literature, where the radio-emission is compared with the optical brightness \citep{kellerman1989,padovani1993,lafranca1994}. The selection criterion of \citet{lyu2022} is very conservative, which might have resulted in many AGN that would have satisfied the traditional radio-AGN criteria \citep[see, e.g.,][among many others]{lemaux2014} being missed in the \citet{lyu2022} AGN sample. \citet{lyu2022} find 43 radio-loud AGN. Our AGN sample contains, on average, $\sim$ 4 radio-loud AGN. 
    
    \item \textbf{radio-slope:} Assuming a power-law spectrum ($f_\nu\propto \nu^{\alpha}$) between the observed 3\thinspace GHz and 6\thinspace GHz bands, \citet{lyu2022} calculate the radio slope of any given object. The object with a slope index $\alpha$ of more than -0.5 with a $2\sigma$ significance is considered to be a flat-spectrum radio source, and identified as an AGN. There are 18 AGN in the \citet{lyu2022} catalog identified using this criteria. Our AGN sample does not contain any AGN identified using this radio-slope criteria. 
    
    \item \textbf{opt-spectroscopy:} This type of AGN shows optical spectra with hydrogen broad emission lines or hard line ratios usually caused by AGN-driven gas ionization. These AGN are identified using either of the two criteria: (i) narrow-line AGN presented in \citet{santini2009} identified using ``Baldwin, Phillips \& Terlevich'' (BPT) \citep{baldwin1981} criteria or (ii) broadline AGN presented in \citet{silverman2010a} identified based on FWHM(H$\alpha)>$ 1000-2000 km/s. \citet{lyu2022} select 22 radio-detected and 26 radio-undetected AGN by cross-matching these AGN with the 3D-HST sample. Hence, they report a total of 48 AGN selected based on opt-spectroscopy-based criteria. Our AGN sample contains, on average, $\sim$ 9 AGN selected from this category. 
    
    \item \textbf{opt-SED:}
    These objects are classified as an optical SED AGN as they have $L_{AGN}>10^8 L_\odot$ for their best-fit SED as well as a significant UV/optical excess above the stellar component which is attributed to an optically blue AGN component. \citet{lyu2022} report 99 opt-SED AGN in their catalog. On average, our AGN sample contains $\sim$ 3 opt-SED AGN. 
    \item \textbf{variable:} These objects exhibit a significant photometric variation at any wavelength within either the X-ray or optical wavebands. They are identified based on their higher median absolute deviation \citep[e.g.,][]{pouliasis2019}. \citet{lyu2022} include variable AGN identified in optical by \citet{pouliasis2019} and in X-ray by \citet{young2012} and \citet{ding2018}. There are 111 variable AGN in the \citet{lyu2022} catalog, and our AGN sample contains, on average, $\sim7$ variable AGN. 
\end{enumerate}

 All 901 AGN reported in \citet{lyu2022} are unique systems, i.e., they each satisfy at least one of the nine AGN selection criteria. We note that the AGN categories are not mutually exclusive, i.e., a given system can be identified as an AGN in multiple categories. In their Figure 1, \citet{lyu2022} shows the overlap of sources in various AGN categories. Additionally, it is also worth noting that type II AGN are not specifically selected in their catalog. 

\subsubsection{AGN matching}\label{subsubsec:agn_matching}

To identify AGN within our galaxy sample, we cross-match the 3D HST coordinates of optical/NIR AGN counterparts provided in the \citet{lyu2022} catalog with the coordinates of galaxies in our galaxy sample, using a nearest-neighbor matching radius of 0.5$\arcsec$. For only one case multiple matches were found, and we selected the match that had the most similar redshifts in the C3VO and the \citet{lyu2022} catalogs. We note that we also checked for a presence of a bulk astrometric offset between our galaxy coordinates from \citet{cardamone2010} and the coordinates of the optical/NIR counterparts of \citet{lyu2022} AGN selected from the 3D-HST catalog and found no appreciable offset. Out of 901 AGN reported in the \citet{lyu2022} catalog, 591 AGN are matched to C3VO (MUSYC) sources within a projected separation of 0.5''. The redshift distribution of all \citet{lyu2022} AGN and the ones that are matched with C3VO sources is shown in Figure~\ref{fig:zhist_agn_lyu}. For approximately 25\% of sources that do not have a match in C3VO, the lack of matching seems to be caused by strong deblending of C3VO sources into multiple sources in the 3D-HST catalog. Most of the remaining sources are missed due to the difference in the depth of observations between the two catalogs, the C3VO catalog adopted here being shallower between the two. In spite of its shallowness, we use the C3VO catalog instead of the deeper catalogs, which are only available for parts of the ECDFS field, to prioritize uniformity across the ECDFS field in our VMC mapping process.

\subsection{Estimation of stellar mass using spectral energy distribution fitting} \label{subsec:gal_sed_fitting}

We estimate the properties of galaxies, such as their stellar mass, by utilizing the SED fitting code \textit{LePhare} \citep{arnouts1999,ilbert2006} for the photometry described above included in the \citet{cardamone2010} catalog. For the SED fitting process, we fix the redshift of the galaxy to its spectroscopic redshift  $z_{spec}$ (if available) or photometric redshift $z_{phot}$. The methodology used here is the same as that used in \citet{lefevre2015,tasca2015,lemaux2022}. Briefly, we fit a range of \citet{bruzual2003} synthetic stellar population models to the observed photometry (magnitudes) in different wavebands. These population models were created based on a \citet{chabrier2003} initial mass function (IMF), a range of dust contents and stellar-phase metallicities, and exponentially declining and delayed star-formation histories (SFHs). 
The output of the SED fitting process in \textit{LePhare} includes the marginalized probability distribution function (PDF) of various physical parameters, such as stellar mass. Stellar masses of galaxies were used only as a test for examining the impact of a stellar mass-limited sample on AGN enhancement, and were not used for the final AGN enhancement results presented in this paper.

\subsection{Measurement of local overdensity using VMC-mapping} \label{subsec:gal_env_meas}

\begin{figure}
\includegraphics[scale=0.26]{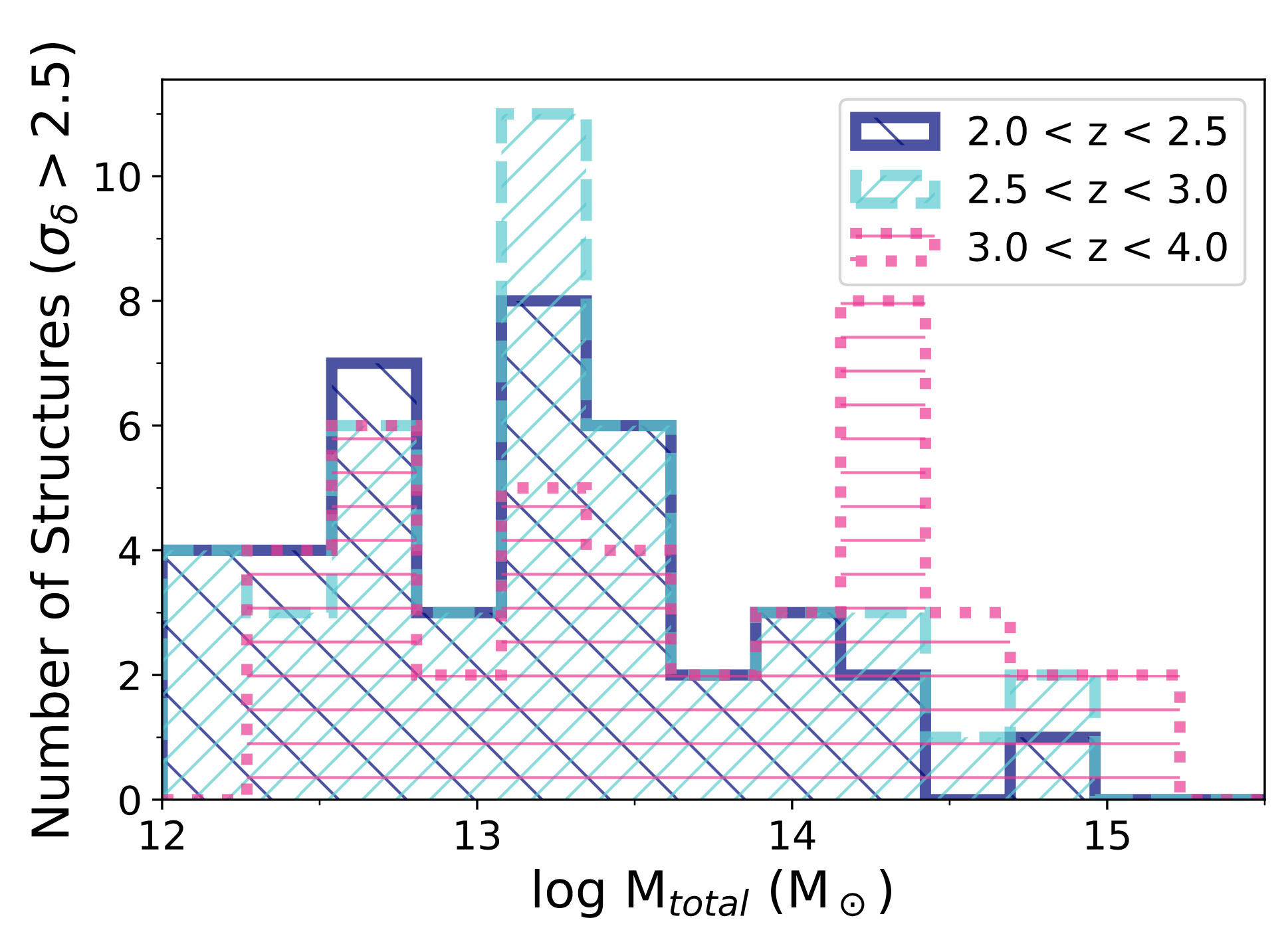}
    \caption{Total mass distribution of massive ($M_{tot}>10^{12}M_\odot$) structures (contiguous envelope consisting of voxels with $\sigma_\delta>2.5$) in three different redshift bins: $2.0<z<2.5$ (navy solid line), $2.5<z<3.0$ (turquoise dashed line), and $3.0<z<4.0$ (pink dotted line).}
    \label{fig:hist_mass_str}
\end{figure}

\begin{figure}
\includegraphics[scale=0.678]{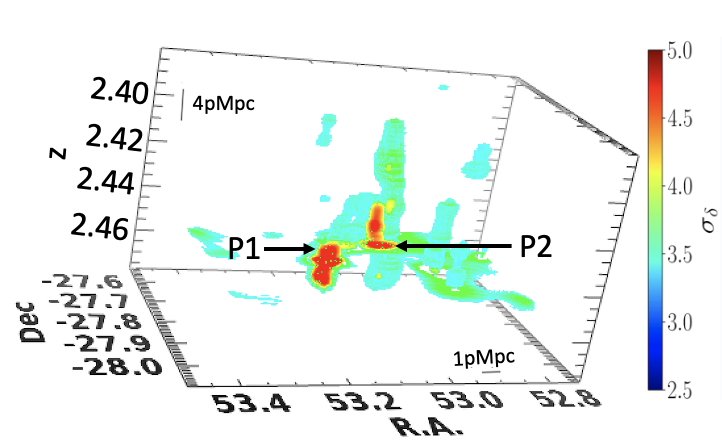}
    \caption{3D distribution of local overdensity in ECDFS, highlighting the most massive protostructure ($M_{\text{tot}}=10^{14.7}M_\odot$) at $z\sim2.45$, spread over $2.40<z<2.48$. The protostructure encapsulates two massive (M$tot>10^{13.25}M_\odot$) overdensity peaks ($\sigma_\delta>5.0$) shown by `P1' and`P2'.  }
    \label{fig:ps_z_24}
\end{figure}

We use a metric of local overdensity $\sigma_\delta$ to measure the environment of galaxies. We calculate this metric using a modified version of the Voronoi tessellation used in many studies \citep[e.g.,][]{lemaux2017,lemaux2018,tomczak2017}. The version of the VMC mapping used for this study is described in detail in \citet{lemaux2022}. This method partitions an area into discrete regions called Voronoi cells. A Voronoi cell consists of all points in the area that are closer to one particular predefined point (galaxy) than to any other predefined points. The edges between these cells are at the same distance from the nearest two or more galaxies, ensuring that each cell is uniquely associated with the closest galaxy. Furthermore, the sizes of these cells vary based on the proximity of galaxies to one another. This variation in cell size effectively captures the variability in galaxy distribution and serves as a robust indicator of local galaxy density.  

This method cannot be directly applied to the redshift dimension due to uncertainties caused by peculiar velocities in spectroscopically confirmed galaxies and relatively large uncertainties in the photometric redshifts. These uncertainties complicate the spatial analysis as they affect the precise location of galaxies in the redshift space. To mitigate these complications, we divide the volume into redshift slices, and apply the VMC method in the projected space of each redshift slice. The redshift slice widths are chosen based on the approximate sizes of protoclusters in simulations \citep{chiang2013,muldrew2015,contini2016,lim2024}, with an additional margin to account for peculiar motion.

Our approach incorporates both spectroscopic and photometric redshifts, adjusting for their respective uncertainties to determine the most suitable redshifts for various Monte Carlo iterations as described in detail in \citet{lemaux2022}. We note that we do not use the spectroscopic redshifts of galaxies as absolute truth, instead we treat them in a probabilistic manner. Similarly, for the probabilistic consideration of the photometric redshifts, we use the median redshift and 16$^{th}$ and 84$^{th}$ percentile of the redshift PDF. Our treatment of the redshifts of galaxies based on Monte Carlo technique is identical to the technique used in \citet{shah24,forrest2023,staab2023} and described in detail in Appendix of \citet{lemaux2022}. Briefly, we generate a suite of 100 Monte Carlo (MC) realizations of $z_{gal}$. If the galaxy does not have a spectroscopic redshift, then $z_{gal}$ is chosen from an asymmetric Gaussian distribution of $z_{phot}$ based on the median value of the photometric redshift and its $1\sigma$ errors (16$^{th}$ and 84$^{th}$ percentile of the photometric redshift PDF). If the galaxy has a spectroscopic redshift, then the selection of $z_{spec}$ as $z_{gal}$ is based on reliability, i.e., the quality flag of the $z_{spec}$. Galaxies with the spectroscopic redshift quality flag of 3 or 4 had their $z_{spec}$ selected as $z_{gal}$ for $\sim99.3\%$ of all 100 MC iterations. For the rest of the iterations, their $z_{gal}$ was drawn from their asymmetric Gaussian distribution of $z_{phot}$. Similarly, for galaxies with the quality flag of 2 or 9, for $\sim70\%$ iterations, $z_{spec}$ was used as $z_{gal}$, and for the rest of the iterations, $z_{gal}$ was drawn from their asymmetric Gaussian distribution of $z_{phot}$. All 100 MC iterations were utilized for the AGN fraction analysis as described in Section~\ref{sec:agn_frac_analysis}.

 The VMC technique yields measurements of galaxy overdensity ($\delta_{gal}$) and the statistical significance of these overdensities ($\sigma_{\delta}$) across a three-dimensional grid aligned with the right ascension (RA), declination (DEC), and redshift (z) axes. For additional information on the computation of these metrics, refer to \citet{forrest2023} and \citet{shah24}. The overdensity values assigned to any given galaxy would correspond to the $\sigma_\delta$ of the nearest Voronoi cell to the galaxy's coordinates and its $z_{gal}$. For 120,525 cases (i.e., average $\sim$ 1205 objects per
MC iteration) where $z_{gal}$ is between $2.0<z_{gal}<4.0$, the redshift vs. $\sigma_\delta$ distribution over all 100 MC iterations is shown in Figure~\ref{fig:z_sig_scat}. As the figure shows, we divide the galaxy sample in 3 different overdensity  bins of (i) $\sigma_{\delta}<2$ (coeval field), (ii) $2<\sigma_{\delta}<5$ (intermediate overdensity), (iii) $\sigma_{\delta}>5$ (highest overdensity peaks) for our analysis (see details in \S\ref{sec:agn_frac_analysis}).

 Using the VMC maps, a protostructure is defined as a contiguous envelope of VMC cells, where each cell has an overdensity of more than $2.5\times$ the RMS of the density distribution in a given slice as measured by a 5$^{th}$ order polynomial to $\sigma$ vs. z. In \citet{shah24}, we presented six spectrosocopically confirmed massive ($M_{tot}\ge10^{14.8}M_\odot$) protostructures at $2.5<z<4.5$. In this study, we exclude the redshift range of $4<z<4.5$ due to limiting depth of multi-wavelength observations significantly affecting the number of AGN (see Figure~\ref{fig:zhist_agn_lyu} and Figure~\ref{fig:sc_lxobs_z_lyu}) in this redshift range. Therefore while the first five protostructures \textit{Drishti} ($M_{tot}\sim10^{14.9}M_\odot$, $z\sim2.67$), \textit{Surabhi} ($M_{tot}\sim10^{14.8}M_\odot$, $z\sim2.80$), \textit{Shrawan} ($M_{tot}\sim10^{15.1}M_\odot$, $z\sim3.3$), \textit{Smruti} ($M_{tot}\sim10^{15.1}M_\odot$, $z\sim3.47$), and \textit{Sparsh} ($M_{tot}\sim10^{14.8}M_\odot$, $z\sim3.70$) from \citet{shah24} are included in this study, the highest redshift structure \textit{Ruchi} ($M_{tot}\sim10^{15.4}M_\odot$) at $z\sim4.14$ is not included. The redshift of these five massive spectroscopically confirmed protostructures is also showed in Figure~\ref{fig:z_sig_scat}.

Furthermore, for this study, we extend the lower bound of the redshift to $z=2$, selecting the redshift range of $2<z<4$ for our analysis. Below this redshift bound ($z<2$), VUDS and C3VO do not effectively target galaxies. By extending the lower redshift bound from $z=2.5$ used in \citet{shah24} to $z=2.0$ in this study, we considerably increase the galaxy sample and the corresponding AGN sample.

The redshift range of $2 < z < 2.5$ was not considered in \citet{shah24} because that study concentrated on six of the most massive protostructures within the redshift range $2.5 < z < 4.5$. These six protostructures are, thus, not coincidentally, the most massive protostructures in the extended range of $2.0 < z < 4.5$, surpassing all protostructures in the $2 < z < 2.5$ range in terms of mass. We show the total mass distribution of all large structures (M$_{tot}>10^{12}$M$_\odot$) in the three redshift bins of $2.0<z<2.5$, $2.5<z<3.0$, and $3.0<z<4.0$ in Figure~\ref{fig:hist_mass_str}. These structures are identified based on a search consistent with the process described in \citet{shah24}. There are considerably more highly massive (M$_{tot}>10^{14}$M$_\odot$) structures at higher redshift compared to in the lowest redshift bin. We present the most massive structure in the redshift range of $2.0<z<2.5$ below.

\subsubsection{Newly detected protostructure at $z\sim2.45$}

The most massive protostructure in the redshift range of $2.0<z<2.5$ is at $z\sim2.45$, which is spread over $2.40<z<2.48$. The 3D distribution of the local overdensity distribution in this protostructure along with its two massive (M$_{tot}>10^{13}$M$_\odot$) overdensity peaks is shown in  Figure~\ref{fig:ps_z_24}. It has a total mass of $M_{tot}=10^{14.7}M_\odot$ and volume of 7387\thinspace cMpc$^3$.

To summarize the galaxy sample selection and its property estimation process described in this entire section, the final galaxy sample consists of all 100 MC iterations as described in \S\ref{subsec:gal_env_meas}. For each galaxy in a given MC iteration, we estimate the stellar mass using SED fitting for its redshift $z_{gal}$, and the overdensity $\sigma_\delta$ value using the VMC maps and the location of galaxy (RA, Dec, $z_{gal}$). Consequently, the redshift $z_{gal}$, and thus the stellar mass and $\sigma_\delta$ of a given galaxy may vary across different MC iterations. We only select galaxies with IRAC1 or IRAC2 magnitudes brighter than 24.8 and $2<z_{gal}<4$. Using these criteria, we generate a sample of 5,514,700 objects (i.e., $\times$ 100 Monte Carlo iterations for 55147 unique objects). Out of these, there are 120,525 cases where $2 < z_{\text{gal}} < 4$ over all 100 iterations. The z$_{\text{gal}}$ and $\sigma_\delta$ distribution of the 120,525 cases (i.e., average $\sim1205$ objects per MC iteration) are shown in Figure~\ref{fig:z_sig_scat}.

\section{AGN Fraction Analysis}\label{sec:agn_frac_analysis}

We define the AGN fraction as the ratio of the number of AGN to the total number of galaxies. We calculate the AGN fraction in each of the 100 MC iterations and use the median AGN fraction value of all iterations as the final AGN fraction. For the errors on the AGN fraction, we use the 16$^{th}$ and 84$^{th}$ percentiles of the AGN fraction in all MC iterations, added in quadrature with the error on the number of AGN fraction values computed by assuming binomial statistics \citep{cameron2011}.

\subsection{Enhancement in AGN fraction with local environment}\label{subsec:agn_frac_w_local_env}

\begin{figure}
\includegraphics[scale=0.7]{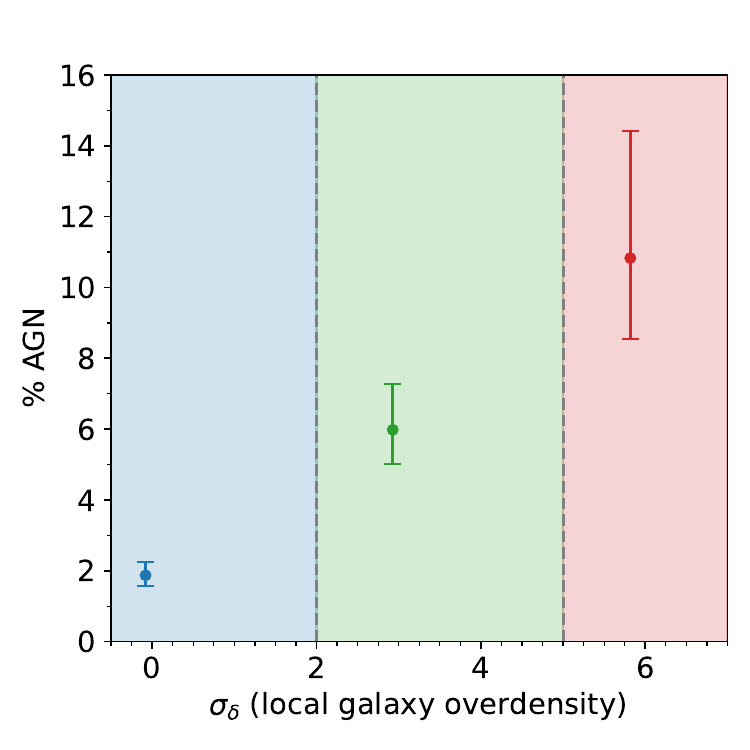}
    \caption{The variation in AGN fraction (in \%) with local overdensity ($\sigma_\delta$) for all different types of AGN combined at $2<z<4$. The $\sigma_\delta$ is divided into three different bins - (i) $\sigma_\delta<2.0$ (blue): coeval field, (ii)  $2.0<\sigma_\delta<5.0$ (green): intermediate overdensity, (iii) $\sigma_{\delta}>5$ (red): highest overdensity peaks. The figure shows a clear trend of increasing AGN fraction with increasing local overdensities of galaxies.} 
    \label{fig:agn_frac_allz_comb}
\end{figure}
\begin{figure*}
\includegraphics[scale=0.6]{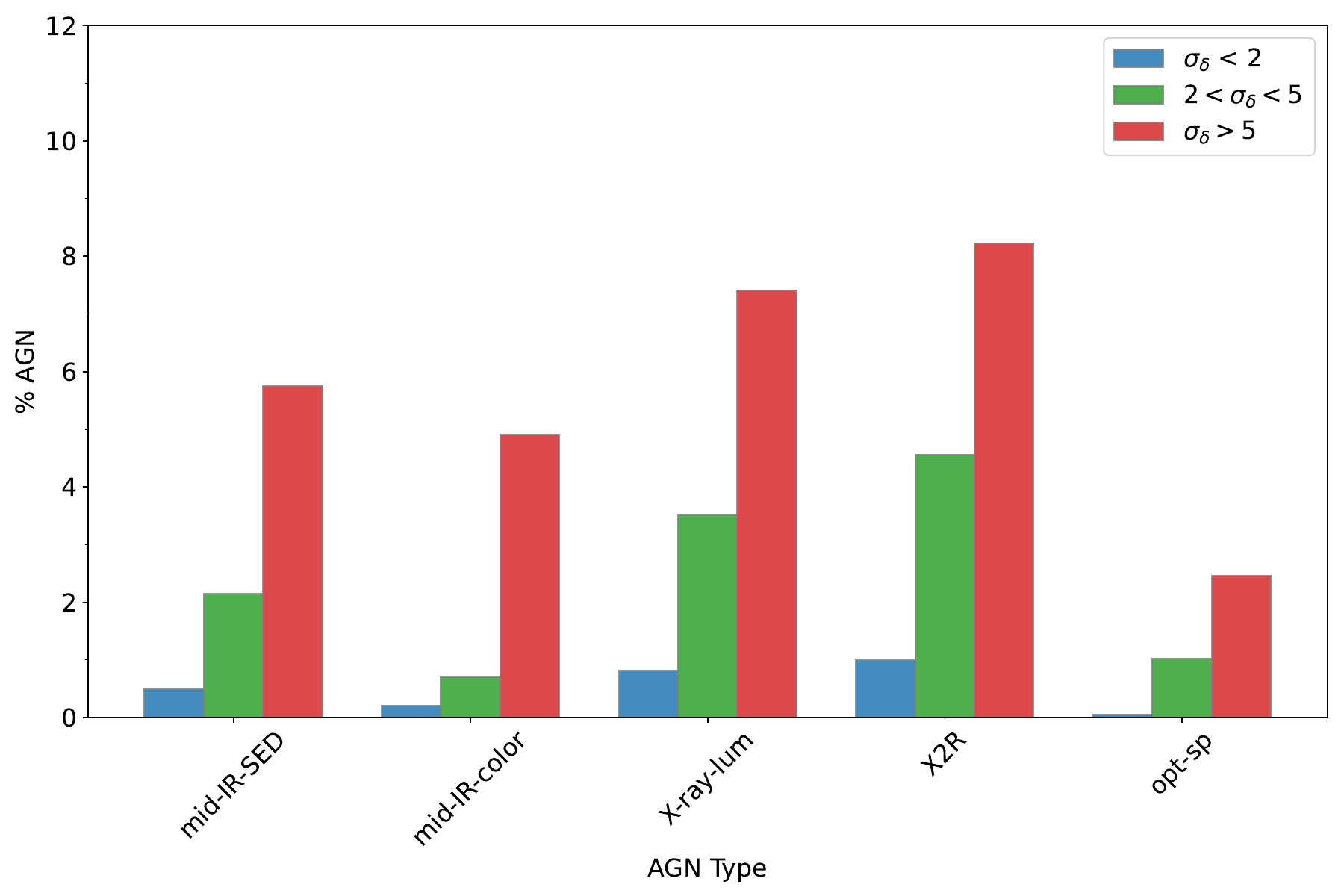}
    \caption{AGN fraction (in \%) for the different types of AGN in different overdensity bins: coeval field (blue)  intermediate overdensity (green), and highest overdensity peaks (red). Here, we only show the five categories (mid-IR-SED, mid-IR-color, X-ray-lum, X2R, and opt-sp) for which, on average, there are more than eight AGN across all 100 MC iterations. Note that the categories are not mutually exclusive; an AGN may meet the criteria for more than one category. See Figure 11 in \citet{lyu2022} for the overlap of sources in different AGN categories. The mean error on these AGN fractions above are $\sim$35\% of their respective values (i.e., $\sim$0.1\% - $\sim$2.5\%). For all categories, there is a clear trend of higher AGN fraction for galaxies in denser environments. } 
    \label{fig:hist_agn_frac_agn_type_allz_comb}
\end{figure*}

For our entire sample ($2.0<z<4.0)$, we show the change in AGN fraction (in \%) of galaxies with local overdensity ($\sigma_\delta$) in Figure~\ref{fig:agn_frac_allz_comb}. As the AGN fraction increases from 1.9$^{+0.4}_{-0.3}$\% for galaxies in the coeval field, i.e., $\sigma_\delta<2$ to 10.9$^{+3.6}_{-2.3}$\% for galaxies in highly overdense peaks, i.e., $\sigma_\delta>5$, which is a clear trend of increasing AGN fraction with increasing overdensity of galaxies. This difference in AGN fraction is at $\sim 3.9\sigma$ level. 

To check the sensitivity of our results to the stellar mass limit of galaxies in our redshift range, we conducted a test to see if the AGN fraction results vary when using a redshift-based 80\% stellar-mass completeness limited galaxy sample instead of our entire sample. The trend described above remains unchanged and we do not see a significant difference in the results for this stellar-mass limited sample as shown in Appendix~\ref{app:1}.

\subsubsection{AGN fraction of various AGN types in different environments}

We show the values of AGN fraction (in \%) corresponding to the different types of AGN in these local overdensity bins in Figure~\ref{fig:hist_agn_frac_agn_type_allz_comb}. There are five categories of AGN that have on average, more than eight AGN across all MC iterations. These categories are: mid-IR, mid-IR-color, X-ray-lum, X2R, and opt-sp. We find that for all of these five categories, there is a clear trend of increasing AGN fraction with increasing overdensity as shown in the figure. Out of all of these categories, we see the highest AGN fraction for the X2R category compared to the rest of the four categories in all environment overdensity bins. For the X2R category, the AGN fraction increases from 1.0$^{+0.5}_{-0.3}$\% to 8.2$^{+3.4}_{-2.0}$ ($\sim 3.5\sigma$ level) as the local ovendensity increases from $\sigma_\delta<2.0$ to $\sigma_\delta>5.0$.

\begin{figure}
\includegraphics[scale=0.7]{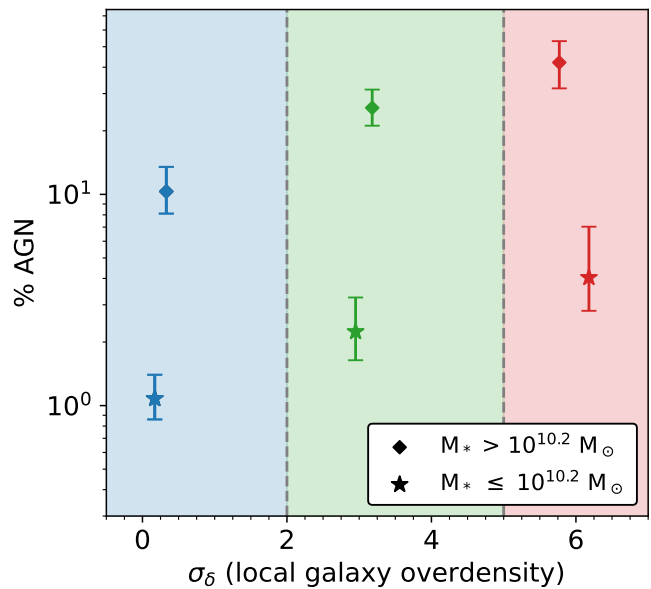}
    \caption{The variation in AGN fraction (in \%) with local overdensity ($\sigma_\delta$) for all different types of AGN combined at $2<z<4$ in two different mass bins divided at $M_*=10^{10.2} M_\odot$. The $\sigma_\delta$ bins are the same as in Figure~\ref{fig:agn_frac_allz_comb}. The diamond points correspond to galaxies with $M_*>10^{10.2} M_\odot$ and the star points correspond to galaxies with $M_*<10^{10.2} M_\odot$. The figure shows a clear trend of increasing AGN fraction with increasing local overdensities of galaxies in both stellar mass bins.}
    \label{fig:agn_frac_mbins_allz}
\end{figure}

\begin{figure}
\includegraphics[scale=0.7]{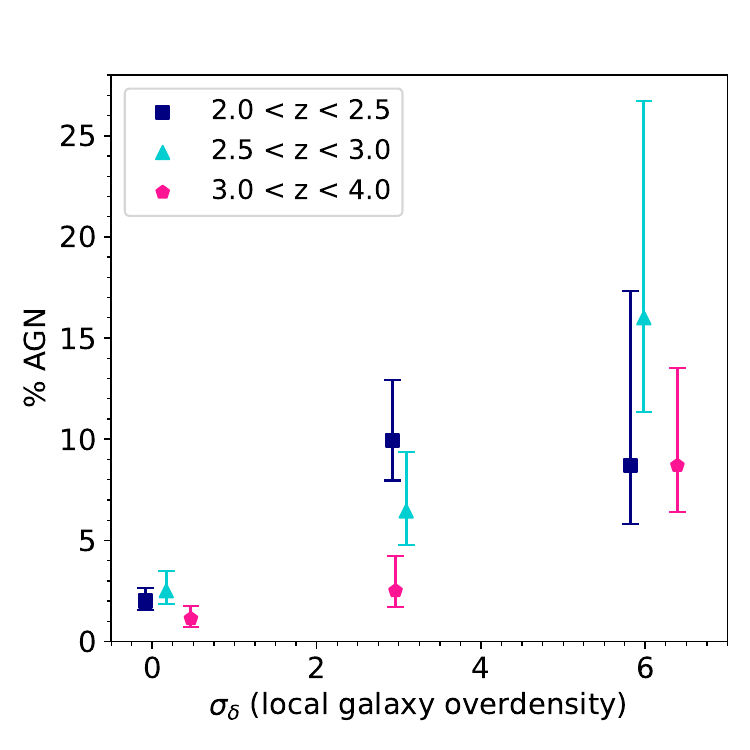}
    \caption{The variation of AGN fraction (in \%) with local overdensity for all different types of AGN combined in three different redshift bins of $2.0<z<2.5$ (navy squares), $2.5<z<3.0$ (cyan triangles), and $3.0<z<4.0$ (pink pentagons). Almost all points show a higher AGN fraction in denser environments at all redshifts. } 
    \label{fig:agn_frac_zdiv}
\end{figure}

\subsubsection{AGN fraction as a function of stellar mass and environment}

To study the change in AGN fraction with environment for different stellar masses of the AGN host galaxies, we divide our entire galaxy sample into two different stellar mass bins: $M_*>10^{10.2} M_\odot$ and $M_*\leq10^{10.2} M_\odot$. The median stellar mass in these two bins differ approximately by an order of magnitude ($\sim10^{9.5}$M$_\odot$ vs. $\sim10^{10.4}$M$_\odot$), however the median stellar mass in different environments is similar (difference $\le 0.1dex$) within each of these two mass bins. The change AGN fraction with the local overdensity in the subsamples of the two mass bins is shown in Figure~\ref{fig:agn_frac_mbins_allz}. For the higher stellar mass bin ($M_*>10^{10.2} M_\odot$), the AGN fraction increases from 10.3$^{+3.2}_{-2.2}$ for $\sigma_\delta<2.0$ to 42.1$^{+11.0}_{-10.4}$ (a $\sim 2.9\sigma$ increment) for $\sigma_\delta<5.0$. Similarly, for the lower stellar mass bin ($M_*<10^{10.2} M_\odot$), the AGN fraction increases from 1.1$^{+0.3}_{-0.2}$ for $\sigma_\delta<2.0$ to 4.0$^{+3.0}_{-1.2}$ (a $\sim 2.3\sigma$ increment) for $\sigma_\delta<5.0$. Notably, at the same local environment for all environmental bins, galaxies in the higher stellar mass bin have an AGN fraction that is $\sim10\times$ higher than their counterparts in the lower stellar mass bins in all three environment bins. This increment ($\sim10\times$) in the AGN fraction with a $10\times$ increase in stellar mass in a given local overdensity environment, is higher compared to the increment ($\sim4\times$) in the AGN fraction with the change in the the local overdensity of environment ($\sigma_\delta<2.0$ to $\sigma_\delta>5.0$) at a given stellar mass of the AGN host galaxies.

\subsubsection{AGN fraction as a function of redshift and environment}

We further divide our entire sample in three different redshift bins of $2.0<z<2.5$, $2.5<z<3.0$, and $3.0<z<4.0$. We show the AGN fraction for different environments in these three redshift bins in Figure~\ref{fig:agn_frac_zdiv}. For $3.0<z<4.0$, the AGN fraction increases from 1.1$^{+0.7}_{-0.4}$\% for $\sigma_\delta<2.0$ to 8.7$^{+4.8}_{-2.3}$\% ($\sim 3.2 \sigma$ level) for $\sigma_\delta>5.0$. Except for the highest overdensity bin for the lowest redshift bin ($\sigma_\delta>5$ and $2.0<z<2.5$), all other points show a clear and continuous trend of increasing AGN fraction with increasing overdensity in all three redshift bins. The AGN fraction in the highest redshift bin is lower than the other two redshift bins, which is likely due to the considerable variation in the completeness of $Lx,obs$ (Figure~\ref{fig:sc_lxobs_z_lyu}) and other multi-wavelength observations used for AGN identification. We also note that the intrinsic AGN fraction for different types can also vary significantly with redshift. So our results at different redshifts are affected by both of these factors. However, even considering these variations, we still see higher AGN fractions in denser local environments compared to the coeval fields at all redshifts.

\subsection{Enhancement in AGN fraction with global environment}

\begin{figure}
\includegraphics[scale=0.263]{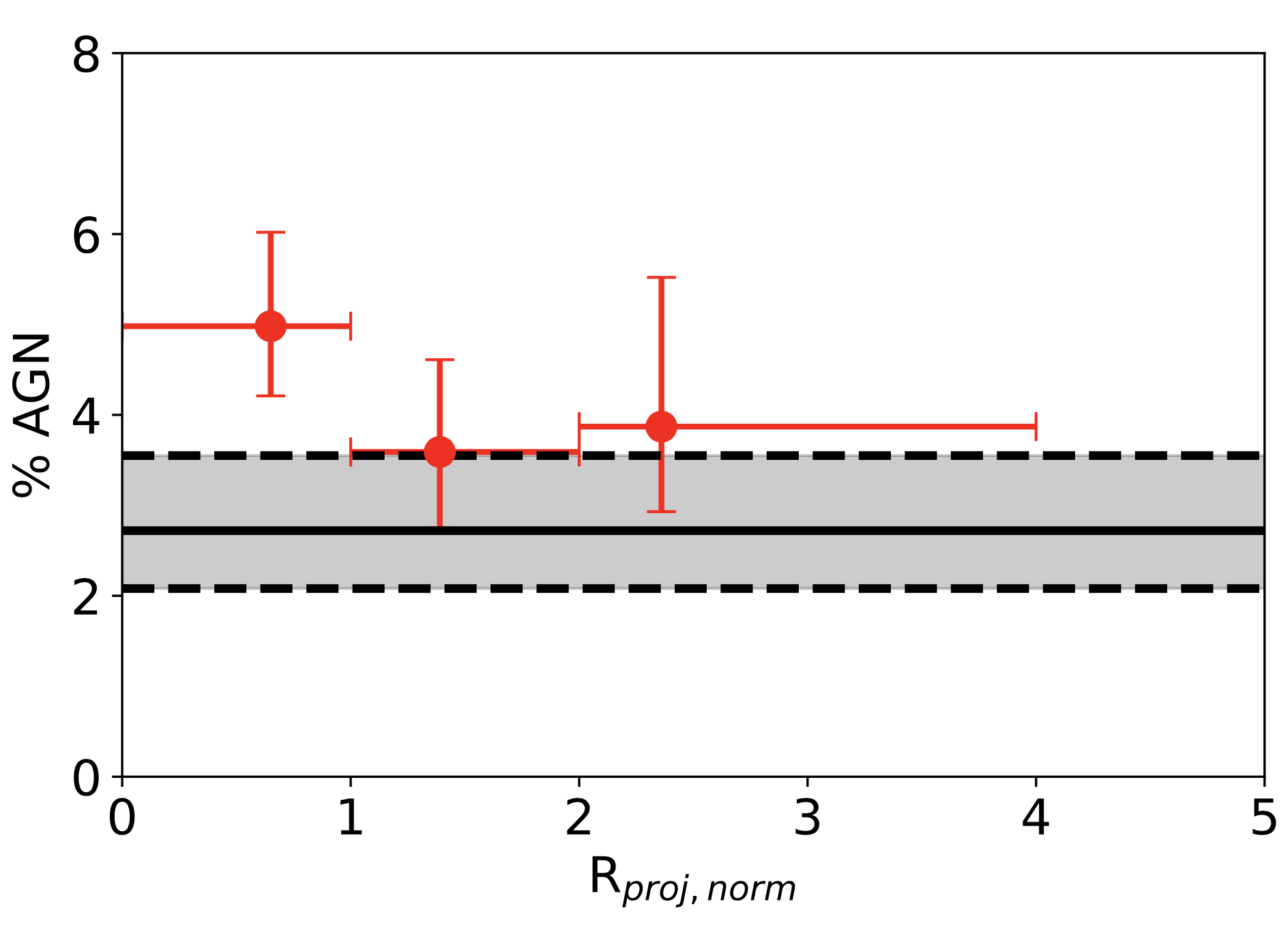}
    \caption{AGN fraction (in \%) as a function of  global environment $R_{proj,norm}=R_{proj}/R_{eff}$ (red points), where R$_{proj}$ is the projected distance of the galaxy from its nearest 5\thinspace $\sigma_\delta$ peak and R$_{eff}$ is the effective radius of the corresponding 5\thinspace $\sigma_\delta$ peak. The solid black line is the AGN fraction value for the corresponding coeval field, with its 1$\sigma$ errors shown using dashed black lines.}
    \label{fig:agn_frac_vs_r_rnorm}
\end{figure}

In addition to local environment metrics, considering the global environment gives a complementary view of the potentially environment-related factors influencing AGN activity. The global environment includes large-scale structures like clusters, filaments, and voids, which can significantly affect galaxy properties and evolution. While local overdensity metrics capturing the immediate surroundings of a galaxy, global overdensity metrics provide insights into its larger scale environment.  The global environment can reveal the influence of large-scale gravitational potential wells and other dynamical processes that might not be apparent from local densities alone. This is particularly important for understanding the role of massive protostructures in galaxy evolution. Furthermore, local and global environments can influence galaxies in different ways. While local density likely correlates with immediate processes like galaxy mergers and interactions, the global environment can impact broader phenomena such as gas accretion, stripping, and the infall of galaxies into larger structures. By combining both local and global measures, we can identify if certain trends in AGN activity are consistent across different scales or if they exhibit scale-dependent behavior. This helps in understanding the multi-scale nature of environmental effects on galaxies.

To study the variation in the AGN fraction of galaxies with their global environment, we compute an environment metric based on the location of galaxies with regards to its closest massive ($M_{tot}>10^{12.8}M_\odot$) $5\thinspace \sigma_\delta$ peak. We define this metric by $R_{proj,norm}=R_{proj}/R_{eff}$. Here, R$_{proj}$ is the projected distance of the galaxy from its closest $5\sigma_\delta$ peak and R$_{eff}$ is the effective radius (R$_{eff}$ = (R$_{x}$+R$_{y}$)/2) of the corresponding $5\sigma_\delta$ peak. For each galaxy in any MC iteration, we first identify all $5\sigma_\delta$ peaks in the ECDFS field where the galaxy's redshift falls within the redshift range of the peak, allowing a buffer of ±0.05 (i.e., z${ext} \pm 0.05$). Additionally, we ensure that the galaxy is within a projected separation (R${proj}$) of less than 10 cMpc from the peak. Then we select the peak for which $R_{proj,norm}$ value is the lowest for the given galaxy in order to study the relation of AGN fraction of galaxies with their global environment. The galaxies that have an associated peak identified using this method are considered to be in dense environments. The galaxies in $2<z<4$ that (i) do not have an associated peak, i.e, $z_{gal}$ outside z$_{ext}\pm0.05$ of all peaks, or (ii) R$_{proj}$ of more than 10\thinspace cMpc - make a coeval field sample used for a comparison. For this analysis, we only consider massive ($M_{tot}>10^{12.8}M_\odot$) $5\sigma$ peaks as they are likely associated to massive protostructures, providing probes for denser large-scale global environments.

The AGN fraction for galaxies with an associated peak as well as galaxies in this coeval field sample is shown as a function of $R_{proj,norm}$ in Figure~\ref{fig:agn_frac_vs_r_rnorm}. For the galaxies closest to the overdense peaks ($R_{proj,norm}<1.0$), i.e., highest global overdensity, we observe a higher AGN fraction of 5.0$^{+1.0}_{-0.8}$\% (a $\sim 2.0\sigma$ increment), compared to AGN fraction (2.7$^{+0.8}_{-0.6}$\%) of the coeval field galaxies. Hence, similar to the local environment result, we see a higher AGN activity for galaxies in a denser environment as compared to coeval field galaxies. Our results are not sensitive to the threshold ($R_{proj,norm}<1$) used for the lowest bin, the mass-threshold $(M_{tot>10^{12.8}M_\odot})$, or the buffer (z$_{ext}\pm0.05$) around the $5\sigma$ peaks. We note that the value of the AGN fraction increases further (though not statistically significantly) for the lowest $R_{proj,norm}$ bin as we decrease the upper limit from $R_{proj,norm}=1$.

The AGN fraction of 5.0$^{+1.0}_{-0.8}$\% for the highest global overdensity ($R_{proj,norm}<1.0$)  is lower compared to the AGN fraction of 10.9$^{+3.6}_{-2.3}$\% for the highest local overdensity ($\sigma_\delta>5.0$) bin shown in Figure~\ref{fig:agn_frac_allz_comb}. Furthermore, the AGN fraction (2.7$^{+0.8}_{-0.6}$\%) of the coeval field defined based on global environment as described above, is higher compared to the AGN fraction (1.9$^{+0.4}_{-0.3}$\%) corresponding to the coeval field defined based on local overdensity   ($\sigma_\delta<2.0$). In other words, the increment in the AGN fraction with changes in local overdensity of galaxies is higher compared to the changes in their global overdensities. Part of this difference is caused by the change in the overdense galaxy sample and coeval field sample selected based on global overdensity compared to local overdensity. As we adopt a redshift cylinder around the peaks to match galaxies with peaks for the global environment measure, galaxies that have a relatively lower local density can end up having relatively higher global overdensity. The resultant may lower the dilution of the contrast in AGN fraction with overdensity, lowering the AGN fraction for the highest global density sample. Similarly, some galaxies that resid in highly locally overdense regions may not live in a globally rich environment. Thus, while global environment provides a complementary view to local environment, the considerate uncertainties associated with measuring and characterizing the latter prevents us from drawing strong conclusions.


\section{Discussion}\label{sec:discussion}

To study the role of environment on the AGN activity of galaxies at high redshift, we conduct an analysis showing the AGN fraction of galaxies in a range of environments - from coeval fields to highly dense protostructure peaks at $2<z<4$. For the combined AGN fraction of all nine AGN types, we see a clear trend of increasing AGN fraction with increasing local overdensity of galaxies. The trend is also present for all five types of AGN that have on average more than eight AGN across all MC iterations, including AGN identified using mid-IR-SED, mid-IR-color, X-ray luminosity, X-ray to radio luminosity ratio, or optical spectroscopy. 

Our trend of increasing AGN fraction with increasing local overdensity of galaxies is in contrast with the local Universe ($z\sim0$), in which studies show lower AGN fractions of luminous AGN in cluster galaxies compared to coeval field galaxies \citep{kauffman2004,popesso2004}. Our trend is also considerably different from studies at $1<z<1.5$, which show similar X-ray and MIR-selected AGN fractions in cluster galaxies and field galaxies \citep{martini2013}.  Our results, combined with these results, suggest a reversal of the AGN-environment relation at high redshift of $z>2$.

As we observe this trend across all three redshift bins in $2<z<4$, including $2.0<z<2.5$, it suggests that the reversal might be occurring before $z\sim2.0$. This result is consistent with studies based on individual protoclusters, such as \citet{krishnan2017} (X-ray AGN; $z\sim1.7$),  \citet{tozzi2022} (X-ray AGN; $z\sim2.156$), \citet{polletta2021} (X-ray AGN; $z\sim2.16$), \citet{digby2010} (AGN identification using emission-lines in optical/near-IR spectra; $z\sim2.2$), and \citet{monson2023} (X-ray AGN; $z\sim3.09$),  all of which show a higher AGN fraction in protocluster galaxies as compared to field galaxies. Additionally, our X-ray-luminosity AGN fraction in the intermediate and highly overdense local overdensity environments at $2<z<4$ are consistent with the X-ray AGN fraction in individual clusters reported in \citet{digby2010} ($z\sim$2.2) and \citet{lehmer2009} ($z\sim$3.09). The enhancement of $\sim2.5\times$ at $3.0<z<4.0$ in the intermediate overdensity bin compared to the field observed in our sample, is within error bars of the enhancement seen in individual protocluster at $z\sim3.09$ in \citet{lehmer2009} (Lyman break galaxies, X-ray AGN sample). However, our findings are different compared to the findings of \citet{macuga2019}, who did not find a relation between environment and X-ray AGN fraction in a $z\sim2.53$ protocluster. 

One possible reason for this difference could be the variation in the dynamical states of protoclusters, which may result in different impacts of the environment on AGN activity in various protoclusters. As a given protocluster can show a range of impacts on AGN activity, a statistical sample like the one utilized in this study is necessary to understand the overall impact of the protocluster population on AGN activity in galaxies. Furthermore, there are also differences in the methods used to characterize protoclusters and their member galaxies between our study and these studies, which may result in variations in AGN fraction enhancement. The substantial photometric and spectroscopic data used for our galaxy sample, and the extensive AGN sample provided by \citet{lyu2022} allow us to have a large sample of galaxies and AGN spanning a wide range of redshift and environment expanding the scope of this type of analysis.

Studies show approximately an order of magnitude increment in the AGN fraction of galaxies with an increment in the host galaxy mass by order of magnitude at lower redshifts($z\le1.0$) \citep[e.g.,][]{juneau2013,aird2021} and in the stellar mass range and redshift range of our study \citep[e.g.,][]{aird2019,guetzoyan2024,martinez2024}. These increments are consistent with our results of approximately an order of magnitude increment in AGN fraction in the higher mass bin compared to the lower mass bin in a given environment as shown in Figure~\ref{fig:agn_frac_mbins_allz}. We note that these studies are based on a single type of AGN (for example, X-ray or optical), as apposed to, the nine different AGN categories included in our study. There are also differences between our study and these studies such as the depth of multi-wavelength observations, methods used to generate galaxy samples, etc. These differences can lead to differences in the AGN fraction values observed in our study compared to these studies.

The higher increments in the AGN fraction with an increase in the stellar mass of the host galaxies in a given local overdensity environment, compared to the increment in the AGN fraction with the change in the local overdensity of galaxies at a given stellar mass, as shown in Figure~\ref{fig:agn_frac_mbins_allz}, suggests that the impact of processes related to the stellar mass-AGN connection is larger compared to that of the environment-AGN connection. As we observe approximately the same increment in the AGN fraction with environment in both higher and lower stellar bins (also shown in Figure~\ref{fig:agn_frac_mbins_allz}), which may suggest that the processes that are responsible for the increment in AGN with environment may not have a strong dependence on the stellar mass of host galaxies.

For a global environment measure $R_{proj,norm}$, our results show signs of increasing AGN fraction with decreasing $R_{proj,norm}$, i.e., decreasing normalized distance from highly overdense ($\sigma_\delta>5$) peaks. Therefore, this result also suggests higher AGN fraction in galaxies in denser global environments. This finding is a contrast of the results of studies in local Universe showing significant decrease in the fraction of X-ray AGN with decreasing cluster-centric radius, going from $r_{500}$ to central regions of clusters \citep[e.g.,][]{ehlert2013, pentericci2013, koulouridis2024}. Our result is also different from the \citet{rambaugh2017} study at $0.65<z<1.28$, showing an absence of a strong relation between AGN activity and location within the large-scale structures (LSSs) in the ORELSE survey \citep{lubin2009}. Therefore, similar to the redshift-based change in the impact of local environment on AGN activity, there seems to be a reversal of the global environment impact on AGN activity at high redshift.

Both of our results, based on local and global environments, show a higher AGN fraction in denser environments compared to coeval-field environments. Our analysis indicates that the enhancement in AGN fraction is more pronounced in the local environment compared to the global environment. This suggests that the processes driving the increase in AGN fraction are more effective on smaller scales rather than larger scales. Smaller-scale processes like galaxy interactions or mergers could be significant factors in this enhancement. Such processes are more common in denser global environments, the global environments can also contribute to the increased AGN fraction observed in these environments, as demonstrated in this study.

At intermediate redshifts and relatively smaller scales, i.e., in spectroscopic galaxy pairs at $0.5<z<3.0$, \citet{shah2020} find no significant enhancement in X-ray AGN or IR-AGN fractions compared to a stellar mass-, redshift-, and environment-matched control sample of isolated galaxies. Similarly, at intermediate redshifts and relatively smaller scales, \citet{martini2013} show that X-ray and MIR-AGN fractions for galaxies in clusters at $1<z<1.5$ are comparable to those of field galaxies. As discussed in \citet{fensch2017}, \citet{shah2020}, and \citet{shah24} (among others), despite the larger gas fraction of galaxies at these intermediate redshifts compared to local galaxies, extreme gas properties (such as high turbulence and temperature) and other processes at these redshifts appear to weaken the nuclear infall of gas that triggers AGN during galaxy intractions and mergers.

In contrast, our study shows that at even higher redshifts of $2<z<5$, the role of the local environment (likely galaxy interactions and mergers) is reversed, as we observe a higher AGN fraction in denser local environments. Several factors could contribute to this observed reversal, such as: the increased gas supply in high-redshift galaxies \citep[e.g.,][]{govardhan2019}, higher merger rates \citep[e.g.,][]{ferreira2020}, and differences in galaxy properties such as stellar mass \citep[e.g.,][]{ownsworth2014} and morphology \citep[e.g.,][]{kartaltepe2013} at these epochs. Our finding suggests a significant redshift evolution in the role of processes influencing AGN activity, highlighting the need for further investigation into how these processes vary across different environment scales and cosmic epochs.

\section{Summary}\label{sec:summary}

We present a study on the relation of the environment of galaxies with its AGN activity at $2.0<z<4.0$. We conduct a robust analysis on the change in AGN fraction of galaxies with the change of local overdensity as well as global overdensity of galaxies. These overdensities are measured using a novel environment measurement technique, utilizing deep multi-wavelength spectroscopic and photometric observations in the GOODS-S field. The AGN in our sample are sourced from the multi-wavelength AGN catalog over nine categories provided by \citet{lyu2022}. We summarize our findings below: 

\begin{enumerate}
    \item For our entire galaxy sample and AGN sample at $2.0<z<4.0$, we see a clear trend of increasing AGN fraction with increasing local overdensity of galaxies. The AGN fraction of galaxies increases from 1.9$^{+0.4}_{-0.3}$\% to 10.9$^{+3.6}_{-2.3}$\% 
    (a $\sim 3.9 \sigma$ increment) as the local overdensity of galaxies increase from $\sigma_\delta<2.0$ (coeval field) to $\sigma_\delta>5.0$ (highly overdense peaks). Our results exhibit $\sim 3\times$ and $\sim5\times$ increment in the AGN fraction compared to the coeval-field for intermediate overdensity bin ($2<\sigma_\delta<5.0$) and highly overdense bin ($\sigma_\delta>5.0$), respectively. 
    \item The trend of increasing AGN fraction with increasing local overdensity of galaxies is present in all five categories of AGN (for which on average there are more than eight AGN across all 100 MC iterations) including mid-IR-SED, mid-IR-color, X-ray luminosity, X-ray-to-radio (X2R) luminosity ratio, and optical-spectroscopy at $2.0<z<4.0$. Note that these categories are not mutually exclusive, i.e., a given system can be identified as an AGN for more than one AGN categories. The highest AGN fraction is 8.2$^{+3.4}_{-2.0}$
    for $\sigma_\delta>5$ for X2R AGN category, which is $\sim 3.5\sigma$ higher than that for the coeval field $\sigma_\delta<2$ galaxies (1.0$^{+0.5}_{-0.3}$).
    \item We divide our sample into two stellar mass bins: $M_*>10^{10.2} M_\odot$ and $M_*\leq10^{10.2} M_\odot$. We observe a clear trend of higher ($\sim4\times$) AGN fractions in denser local overdensity environments in both stellar mass bins. For a given environment, the AGN fractions of the galaxies in the higher stellar mass bins are $\sim10\times$ higher compared to the galaxies in the lower stellar mass bins in all three environment bins.
    \item We split the sample in 3 different redshift bins: $2.0<z<2.5$, $2.5<z<3.0$, and $3.0<z<4.0$. The results for all 3 redshift bins show the trend of increasing AGN fraction with increasing local overdensity of galaxies, with an exception of $\sigma_\delta>5$ bin for $2.0<z<2.5$, which is likely affected by low-number statistics. 
    For $3.0<z<4.0$, the AGN fraction increases from 1.1$^{+0.7}_{-0.4}$ for $\sigma_\delta<2.0$ to 8.7$^{+4.8}_{-2.3}$ (a $\sim 3.2 \sigma$ increment) for $\sigma_\delta>5.0$. 
    \item For global environment measure R$_{proj,norm}$, we see a higher AGN fraction (5.0$^{+1.0}_{-0.8}$\%) ($\sim 2.0\sigma$ higher) for galaxies in denser global environment (R$_{proj,norm} < 1.0$) compared the AGN fraction (2.7$^{+0.8}_{-0.6}$\%) of the corresponding coeval field galaxies. 
\end{enumerate}

Our results, as compared to cluster studies of the low and intermediate redshift Universe, suggest a reversal in the role of environment in AGN activity. Specifically, at high redshifts of $2.0 < z < 4.0$, galaxies in denser environments exhibit a substantially higher AGN fraction compared to their coeval field counterparts, in contrast to the behavior seen in local clusters. This enhancement in AGN fraction in denser environments at high redshift could be caused by factors such as increased gas supply, higher merger rates of galaxies, and differences in the properties of galaxies or environments. Such prevalence of AGN activity in these denser environments could result in increase AGN-feedback processes (such as AGN-driven outflows), expelling or heating gas, which may lead to quenching of star formation of galaxies \citep[e.g.,][]{xie2024}. Our study provides a pathway to better understand the complex interplay between the environment and AGN activity in galaxies at high redshift, and offers valuable constraints on environment-related impacts on AGN activity for theoretical models.

\section{Data Availability}
The data used this study would be shared based on a reasonable request to the corresponding author.

\begin{acknowledgements}
   Results in this paper were partially based on observations made at Cerro Tololo Inter-American Observatory at NSF’s NOIRLab, which is managed by the Association of Universities for Research in Astronomy (AURA) under a cooperative agreement with the National Science Foundation. Supported by the international Gemini Observatory, a program of NSF NOIRLab, which is managed by the Association of Universities for Research in Astronomy (AURA) under a cooperative agreement with the U.S. National Science Foundation, on behalf of the Gemini partnership of Argentina, Brazil, Canada, Chile, the Republic of Korea, and the United States of America. Results additionally relied on observations collected at the European Organisation for Astronomical Research in the Southern Hemisphere. This work is based on observations made with the Spitzer Space Telescope, which is operated by the Jet Propulsion Laboratory, California Institute of Technology under a contract with NASA. Some of the data presented herein were obtained at Keck Observatory, which is a private 501(c)3 non-profit organization operated as a scientific partnership among the California Institute of Technology, the University of California, and the National Aeronautics and Space Administration. The Observatory was made possible by the generous financial support of the W. M. Keck Foundation. This work was supported by NASA’s Astrophysics Data Analysis Program under grant number 80NSSC21K0986. Some of the material presented in this paper is based upon work supported by the National Science Foundation under Grant No. 1908422.

   The authors wish to recognize and acknowledge the very significant cultural role and reverence that the summit of Maunakea has always had within the indigenous Hawaiian community. We are most fortunate to have the opportunity to conduct observations from this mountain. For the purpose of open access, the author has applied a Creative Commons Attribution (CC BY) license to any Author Accepted Manuscript version arising from this submission.

\end{acknowledgements}

%
\bibliographystyle{aa} 
\bibliography{main} 
%
\begin{appendix} 
\section{AGN fraction in a stellar-mass completeness limited sample} \label{app:1}
\begin{figure}
\includegraphics[scale=0.7]{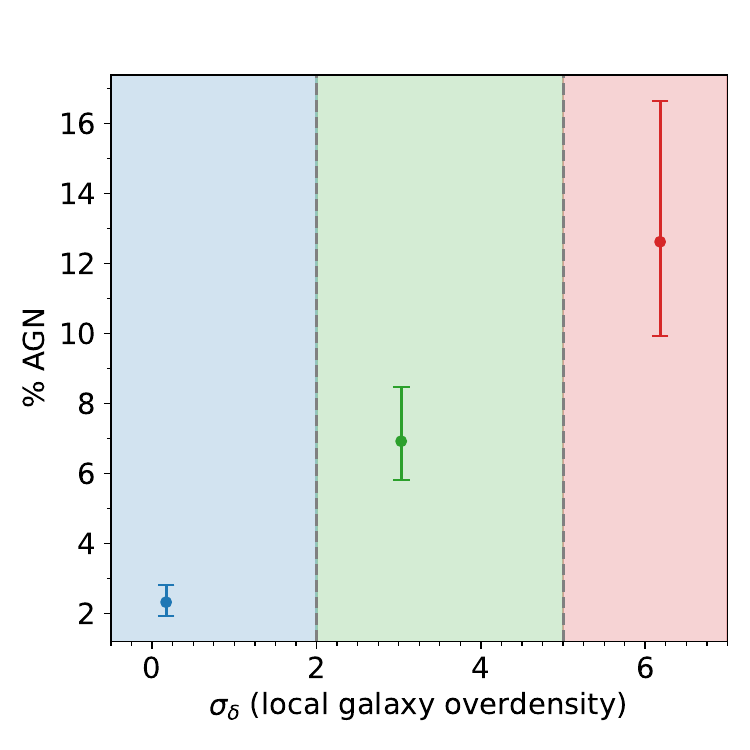}
    \caption{The variation in AGN fraction (in \%) with local overdensity ($\sigma_\delta$) for all different types of AGN combined at $2<z<4$ in an 80\% stellar-mass ($M_*>10^{9.14}M_\odot$) complete in our redshift range. The $\sigma_\delta$ bins are same as in Figure~\ref{fig:agn_frac_allz_comb}. The figure shows a clear trend of increasing AGN fraction with increasing local overdensities of galaxies.} 
    \label{fig:app:agn_frac_mcut_mgt914}
\end{figure}

In the beginning of \S\ref{subsec:agn_frac_w_local_env}, in Figure~\ref{fig:agn_frac_allz_comb}, we show AGN fraction in three different environment bins for our entire galaxy sample in $2.0<z<4.0$. We conduct the same analysis for an 80\% stellar-mass completeness limited galaxy sample ($M_*>10^{9.14}M_\odot$) in our redshift range and present the results in Figure~\ref{fig:app:agn_frac_mcut_mgt914}. The AGN fraction increases from 2.3$^{+0.5}_{-0.4}$\% to 12.6$^{+4.0}_{-2.7}$\% (a $\sim3.8\sigma$ increment) as the local overdensity increases from $\sigma_\delta<2$ (coeval field) to $\sigma_\delta>5$ (overdense peaks). Hence, the AGN fraction values are within error bars and the AGN fraction increment significance is approximate the same when these results are compared to the  results for the overall galaxy sample presented in Figure~\ref{fig:agn_frac_allz_comb}.

\end{appendix}

\end{document}